\documentclass[pre,twocolumn,aps,showpacs,superscriptaddress]{revtex4}
\usepackage{epsfig}
\usepackage{graphics}
\usepackage{dcolumn}
\usepackage{bm}
\usepackage{amssymb}
\usepackage{fancyhdr}

\begin{document}

\title{Orbital ice: an exact Coulomb phase on the diamond lattice}
\author{Gia-Wei Chern}
\affiliation{Department of Physics, University of Wisconsin, Madison, 
WI 53706, USA}

\author{Congjun Wu}
\affiliation{Department of Physics, University of California, San Diego, 
CA 92093, USA}


\begin{abstract}
We demonstrate the existence of orbital Coulomb phase as the exact ground 
state of $p$-orbital exchange Hamiltonian on the diamond lattice. 
The Coulomb phase is an emergent state characterized by algebraic 
dipolar correlations and a gauge structure resulting from local 
constraints (ice~rules) of the underlying lattice models. 
For most ice models on the pyrochlore lattice, these local constraints 
are a direct consequence of minimizing the energy of each individual 
tetrahedron. On the contrary, the orbital ice rules are emergent 
phenomena resulting from the quantum orbital dynamics. 
We show that the orbital ice model exhibits an emergent geometrical
frustration by mapping the degenerate quantum orbital ground states
to the spin-ice states obeying the 2-in-2-out constraints on 
the pyrochlore lattice. We also discuss possible realization of the 
orbital ice model in optical lattices with $p$-band fermionic cold atoms.
\end{abstract}
\maketitle

\section{Introduction}

Common water ice, a strongly correlated proton system, is a canonical 
example of geometrical frustration \cite{ice}. The oxygen ions in ice 
form a periodic diamond lattice whereas the protons are disordered due 
to the frustrated arrangement of two inequivalent O--H bonds with different 
lengths. 
This in turn leads to a macroscopic degeneracy of possible ground states 
and a finite entropy density of ice as temperature tends toward zero. 
Despite being disordered, the positioning of protons dictated by the 
so-called {\it ice rules} exhibits a strong short-range correlation 
in which each oxygen ion has two-near and two-far protons. 
The ice rules also forbid single proton hopping and only allow for 
ring-exchange-type motion, reminiscent of the physics of gauge theory.

A magnetic analogue of ice was discovered in pyrochlore oxides
Dy$_2$Ti$_2$O$_7$ and Ho$_2$Ti$_2$O$_7$ more than a decade ago \cite{spinice}. 
These so-called {\it spin ice} compounds
are essentially pyrochlore Ising magnet in which magnetic moments residing on
a network of corner-sharing tetrahedra are forced by single-ion anisotropy 
to point along the local $\langle 111 \rangle$ axes. 
It is found that extensively degenerate spin configurations obeying the
so-called `2-in-2-out' rules have essentially the same energy over a wide 
range of temperatures. The measured residual entropy is well 
approximated by the Pauling entropy for water ice \cite{ramirez}.
These local constraints require that every tetrahedron 
has two spins pointing in and two pointing out, in apparent analogy with
the ice rules. Reversing a single spin in the ice state creates one
defect tetrahedron with 3-in-1-out spins and another one with 1-in-3-out spins.
As recently pointed out in Ref.~\cite{monopole}, these defect tetrahedra behave exactly as a gas of magnetic 
monopoles interacting with each other via Coulomb's $1/r$ law.

Artificial versions of spin ice have also been created using lithographically 
fabricated arrays of nanoscale magnets \cite{squareice,kagomeice}.
Other proposals of artificial ice systems include charged colloidals in optical
traps and superconducting vortices in specially fabricated pinning 
centers \cite{other-ice,other-ice2}. A valence bond liquid phase with an ice-like
degeneracy is also shown to be the ground state of a spin-1/2 Klein  
model on the pyrochlore lattice \cite{nussinov}.

In most of these ice systems, the fundamental degrees of freedom are doublet
variables defined on the pyrochlore lattice or its two-dimensional counterpart. 
Their Hamiltonians can often be cast into the form
\begin{eqnarray}
	\label{eq:H_ice}
	H_{\rm ice} = J\sum_{\boxtimes} \mathcal{K}_{\boxtimes} + \epsilon\, H'
\end{eqnarray}
where $J > 0$ is the energy scale of excitations and the sum is over
all tetrahedra. $\mathcal{K}_{\boxtimes}$ is a 
nonnegative-definite operator defined for a tetrahedron.
The last term denotes perturbations of energy scale $\epsilon$.
The ice rules correspond to the contraints:
\begin{eqnarray}
	\label{eq:K0}
	\mathcal{K}_{\boxtimes} = 0, 
\end{eqnarray}
for all tetrahedra. Take spin ice as an example, the spin configurations can be specified
by a set of Ising variables $\{\sigma_i\}$ such that $\mathbf S_i = \sigma_i S\,\hat\mathbf e_i$,
where $\hat\mathbf e_i$ denotes the local easy axis. The constraint operator
is then given by $\mathcal{K}_\boxtimes \propto (Q_{\boxtimes})^2$, where
$Q_{\boxtimes} \equiv \sum_{m\in \boxtimes} \sigma_m$ 
is the effective magnetic charge of a tetrahedron. The six up-up-down-down Ising
configurations selected by constraints (\ref{eq:K0}) correspond to the 2-in-2-out rules.
Another example is the spin-$1/2$ Klein model for which $\mathcal{K}_\boxtimes 
\equiv \mathcal{P}_{S_{\boxtimes} = 2}$ is the projection operator onto the subspace 
of maximum total spin $S_\boxtimes = 2$ \cite{nussinov}. 
For temperatures in the regime $\epsilon \ll T \ll J$, 
configurations satisfying the `ice rules' (\ref{eq:K0}) for all tetrahedra,
are effectively degenerate. 

The ice model~(\ref{eq:H_ice}) hosts an emergent Coulomb phase
in which the local constraints $\mathcal{K}_\boxtimes = 0$ translate to a divergence-free
flux $\nabla\cdot\mathbf B = 0$ in the coarse-grained approximation. The effective theory
for the Coulomb phase is equivalent to conventional magnetostatics \cite{henley10}. It follows
that both the `magnetic' field $\mathbf B$ and spins $\mathbf S_i$ in this disordered yet highly
constrained phase exhibit a dipolar-like correlation function $\langle B_{\alpha}(0) 
B_{\beta}(\mathbf r) \rangle \propto \left(\delta_{\alpha\beta} - 3 \hat r_{\alpha} \hat r_{\beta}\right)/r^3$
at large distances.

In this paper, we present an ice model in which the basic degrees
of freedom are triplet orbital variables defined on the {\it diamond} lattice.
Our investigation is partly motivated by recent progress in orbital-related
many-body phenomena in optical lattices. We show that the strong directional dependence
of orbital exchange combined with the special geometry of diamond lattice gives rise to 
a huge degeneracy in the Gutzwiller-type ground states which are also 
exact eigenstates of the orbital exchange Hamiltonian. 
We demonstrate the existence of an orbital Coulomb phase by mapping the degenerate 
orbital manifold to spin-ice states on the medial pyrochlore lattice. 
It is worth noting that the orbital `ice rules' are {\it not} constraints imposed by the Hamiltonian.
Instead, they are emergent properties characterizing the short-range orbital correlations. 
This is in stark contrast to pyrochlore ice models~(\ref{eq:H_ice}) 
in which the ice rules are explicitly incorporated in the Hamiltonian as the minimum energy states of  
individual tetrahedron.

\section{Orbital exchange Hamiltonian}

The ability to precisely control the interaction 
strength of cold atoms in optical lattices provides clean realizations of
strongly correlated models without many undesirable
complexities usually encountered in material systems \cite{review,bloch08}.
In particular, since the cold-atom systems are free of Jahn-Teller 
distortions, they offer a new opportunity to investigate the intrinsic exchange physics associated with 
the orbital degrees of freedom \cite{lewenstein11}. One of the most interesting directions is the novel
frustration phenomenon originating from the anisotropic orbital interactions.

The exchange physics of $p$-orbitals in two-dimensional optical lattices has been
extensively discussed in Refs.~\cite{wu08,zhao08}. 
The intricate interplay between lattice geometry and anisotropic orbital exchange
leads to dramatically distinct ground states in different 
lattices. The orbital exchange on a square lattice is dominated by an 
antiferromagnetic Ising-like Hamiltonian, which gives rise to a N\'eel-type 
orbital order. For triangular, honeycomb, and kagome lattices, the orbital 
interaction is described by a novel quantum 120$^\circ$ model. 
Although long-range orbital orders occur in the cases of triangular and 
kagome lattices, orbital interactions are frustrated on the bipartite 
honeycomb lattice and a huge degeneracy remains in the classical
ground state. These highly degenerate ground states 
can be mapped to fully packed non-intersecting loops on the honeycomb lattice.
Quantum fluctuations, on the other hand, select a six-site plaquette 
ground state through order from disorder mechanism. 
The 120$^\circ$ model also describes the effective orbital interaction 
in transition metal oxides including honeycomb, cubic, and pyrochlore 
lattices \cite{nagano,khomskii,chern10}.

Here we consider a $p$-band Hubbard model with spinless fermions on three-dimensional optical lattices.
We assume that each optical site is approximated by an isotropic harmonic potential.
For two particles per site, one of them fills the inert $s$-orbital while the other one occupies 
one of the three $p$-orbitals. The kinetic terms of $p$-band fermions include a longitudinal 
$t_{\parallel}$ and a transverse $t_{\perp}$ hopping, corresponding 
to $\sigma$ and $\pi$-bondings, respectively.
Typically, $t_{\parallel} \gg t_{\perp}$ \cite{isacsson05} and we shall neglect the transverse 
hopping as a zeroth-order approximation. The fermions interact with each other
through an on-site repulsion: 
$
H_{\rm int} = U\sum_{i, \alpha \neq \beta} n_{i \alpha} n_{i \beta},
$
where $n_{i\,\alpha} = p^\dagger_{i \alpha} p^{\phantom{\dagger}}_{i \alpha}$ 
is the fermion number operator. 
The leading contribution to $U$ comes from the $p$-wave scattering for spinless 
fermions.
The strong correlation regime $U \gg t_{\parallel}$ can be potentially realized with 
the aid of the recently proposed stable optical $p$-wave Feshbach 
resonance \cite{goyal2010}, which has the advantage of suppressing the
high rate of three-body recombination. It should be noted that recent experimental
results have shown some limitations of the optical $s$-wave Feshbach scheme \cite{blatt}.
Further experimental investigations are needed in order to verify the feasibility of
increasing $p$-wave interaction through optical Feshbach resonance.

With charge fluctuations suppressed in the Mott-insulating limit, 
there still remains a triplet orbital degrees of freedom at each site. 
Exchange interactions between these localized orbital variables originate
from the second-order virtual hopping of the fermions. 
Since we assume a dominating $t_{\parallel}$, for a bond parallel to 
$\hat\mathbf n = (n_x, n_y, n_z)$, longitudinal hopping is possible 
only when one of the particles occupies the orbital 
$|\hat\mathbf n\rangle = n_x |p_x\rangle + n_y |p_y\rangle + n_z |p_z\rangle$,
while the other one is in an orthogonal state. 
The energy gain of such an antiferro-orbital alignment is described 
by the Hamiltonian
\begin{eqnarray}
\label{eq:Hex}
H_{\rm ex} = -J \sum_{\langle ij \rangle} \left[ P_i^{\hat\mathbf n_{ij}} 
\bigl({I}-P_j^{\hat\mathbf n_{ij}}\bigr) + 
\bigl({I}-P_i^{\hat\mathbf n_{ij}}\bigr) P_j^{\hat\mathbf n_{ij}}\right].
\end{eqnarray}
Here $J=t^2_{\parallel}/U$ sets the exchange energy scale, ${I}$ is the identity operator, 
and $P^{\hat\mathbf n_{ij}} = |\hat\mathbf n_{ij}\rangle \langle \hat\mathbf n_{ij}|$ 
is the projection operator of the active orbital on a nearest-neighbor bond 
$\langle ij \rangle$.
Obviously, the nature of the orbital exchange physics depends critically 
on the lattice geometry.

\section{Cubic optical lattice} 

\begin{figure}[t]
\includegraphics[width=0.99\columnwidth]{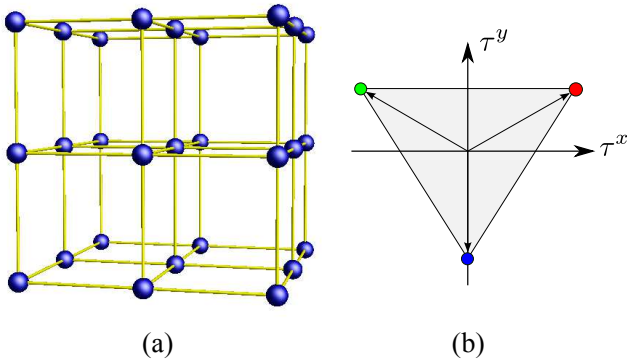}
\caption{(a) Cubic optical lattice. (b) Domain of doublet vector $\langle \bm\tau \rangle$ 
for the cubic lattice. The three corners correspond to $p_x$, $p_y$, and $p_z$ orbitals, respectively.}
\label{fig:domain}
\end{figure}

As a warm-up, we first consider the case of cubic lattice [Fig.~\ref{fig:domain}(a)]. 
Using a basis spanned by $|p_x\rangle$, $|p_y\rangle$, and $|p_z\rangle$ states,
the orbital projectors along the $x$, $y$, and $z$ bonds can be expressed in terms 
of Gell-mann matrices $\lambda^{(3)} = \mbox{diag}(1,-1,0)$ 
and $\lambda^{(8)} = \mbox{diag}(1,1,-2)/\sqrt{3}$. 
By grouping them into a doublet operator $\bm\tau = (\tau^x, \tau^y) 
= (\sqrt{3}/2) (\lambda^{(3)}, \lambda^{(8)})$, the three orbital projectors are  
\begin{eqnarray}
\label{eq:pa}
P^{a} = \left({I} + 2\,\bm\tau\cdot\hat\mathbf e_a\right)/3, 
\quad \quad (a = x, y, z),
\end{eqnarray}
with $\hat\mathbf e_{x/y} = (\pm\frac{\sqrt{3}}{2}, \frac{1}{2})$ 
and $\hat\mathbf e_z = (0,-1)$ [Fig.~\ref{fig:domain}(a)]. 
The expectation value of the doublet vector $\langle \bm\tau\rangle$ represents
the disparities of on-site orbital occupation numbers.
The domain of $\langle \bm\tau \rangle$ is an equilateral triangle [Fig.~\ref{fig:domain}(b)],
whose three corners, $\langle \bm\tau \rangle = \hat\mathbf e_x$, $\hat\mathbf e_y$ and
$\hat\mathbf e_z$, correspond to states with pure $p_x$, $p_y$, and $p_z$ orbitals, respectively.
Substituting the projectors $P^a$ into Eq.~(\ref{eq:Hex}), we obtain an effective 
Hamiltonian:
\begin{eqnarray}
\label{eq:H1}
H_{\rm cubic} =  \frac{8J}{9} \sum_{a = x, y, z}\, 
\sum_{\langle ij \rangle \parallel a}
\left(\bm\tau_i\cdot\hat\mathbf e_a\right)
\left(\bm\tau_j\cdot\hat\mathbf e_a\right),
\end{eqnarray}
up to an irrelevant constant $c_0 = -4NJ/3$. 
Although Eq.~(\ref{eq:H1}) has the same form as the well-known 120$^\circ$
model, it is actually a classical Hamiltonian since 
the three orbital projectors $P^a$ commute with each other.
As a result, the eigenstates of $H_{\rm cubic}$ are simultaneous eigenstates
of the orbital occupation operators $P^{a}_i$ whose eigenvalues are 0 or 1. 
Since each site has exactly one fermion, $P^x + P^y + P^z = 1$, the orbital state
at a given site can be specified by one of the three corners in the triangular domain
of $\langle \bm\tau\rangle$. Eq.~(\ref{eq:H1}) can then be viewed as a 3-state Potts model with anisotropic interactions.
Take an $x$-bond for example, there are 3 different orbital configurations: 
$(p_x, p_x)$, $(p_{y/z}, p_{y/z})$, and $(p_x, p_{y/z})$ whose 
energies are $8J/9$, $\,2J/9$ and $\, -4J/9$, respectively. 

\begin{figure}[t]
\includegraphics[width=0.95\columnwidth]{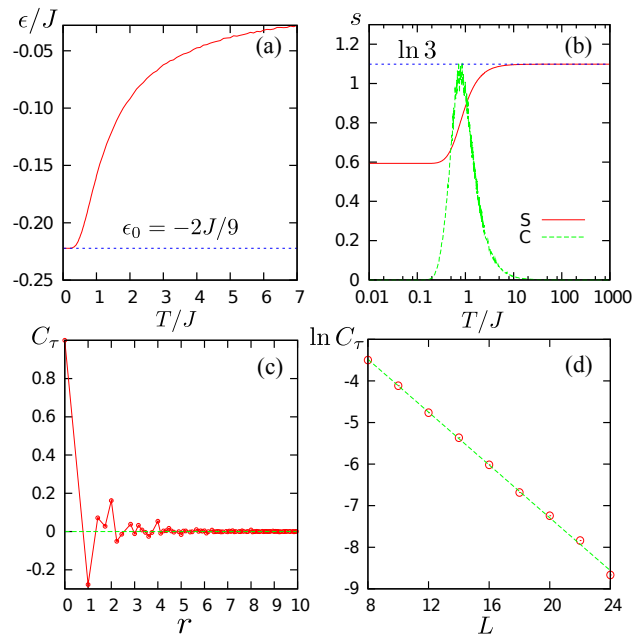}
\caption{Monte Carlo simulations of the classical Hamiltonian~(\ref{eq:H1}).
(a) and (b) show the temperature dependence of average bond energy 
$\epsilon \equiv \langle H_{\rm cubic}\rangle/3N$, specific heat $c$ and 
entropy density $s$, respectively.
The dashed line in (b) indicates the entropy 
density $\ln 3$ at the high-temperature para-orbital phase. 
The orbital correlation function $C_{\tau}(r)=\langle 
\bm\tau(r)\cdot\bm\tau(0)\rangle$ is shown in (c) as 
a function of separation $r$. 
(d) shows $\ln C_{\tau}(L/2)$ as a function of linear system size $L$.}
\label{fig:mc-potts}
\end{figure}

To investigate the orbital correlations in the ground state, we performed 
classical Monte Carlo simulations with periodic boundary conditions on 
systems up to $N=24^3$ sites. 
Figs.~\ref{fig:mc-potts}~(a) and~(b) show the average bond energy $\epsilon$, 
specific heat $c$, and entropy density $s$ as functions of temperature $T$.
The bond energy approaches $\epsilon_0 = -2J/9$ as $T \to 0$, implying that 
2/3 of the bonds with an energy $\epsilon = -4J/9$ are in the antiferro-orbital 
ground states, while the remaining 1/3 are frustrated with an energy of $2J/9$. 
The macroscopic degeneracy of the ground states is evidenced by a residual
entropy density $s_0 \approx 0.599\,k_B$  obtained by integrating
the specific-heat curve [Fig.~\ref{fig:mc-potts}(b)]. The orbital correlation $C_{\tau}(r) = \langle 
\bm\tau(r)\cdot\bm\tau(0)\rangle$ decays rather fast and is negligible beyond 
$r\approx 5$, indicating a disordered orbital liquid. At large separations, 
the correlation function decays exponentially as shown in Fig.~\ref{fig:mc-potts} (c).

The large residual entropy $s_0 \approx 0.599\,k_B$ also implies that the ground state
is susceptible to nominally small perturbations present in the system. Indeed, 
as recently reported in Ref.~\cite{hauke}, inclusion of orbital interactions
which break time-reversal symmetry induces long-range orbital ordering.
As a final remark, it is worth noting that Eq.~(\ref{eq:H1}) is related to but 
quite different from the 120$^\circ$ model with classical O(2) spins,
in which orbital-ordering is shown to be induced via order-from-disorder mechanism
on the cubic lattice \cite{o2-120,trebst10,wenzel11}.

\section{Diamond optical lattice}

\begin{figure}[t]
\includegraphics[width=0.99\columnwidth]{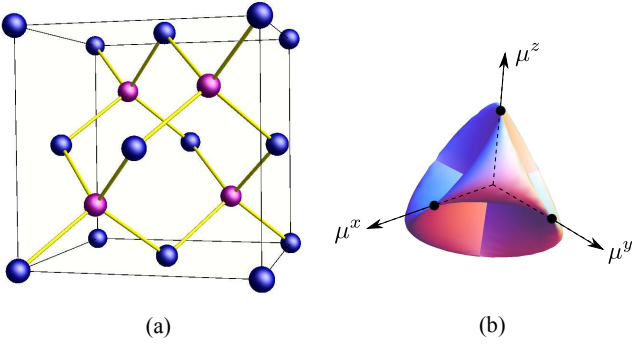}
\caption{(a) Diamond optical lattice. 
(b) Domain of pseudovector $\langle \bm\mu \rangle$ for the diamond lattice.}
\label{fig:domain2}
\end{figure}

We now turn to orbital exchange on the oblique diamond lattice [Fig.~\ref{fig:domain2}(a)].
There are four distinct types of nearest-neighbor bonds pointing along 
directions $\hat\mathbf n_0 = [111]$, $\hat\mathbf n_1 = [1\bar 1\bar 1]$, 
$\hat\mathbf n_2 = [\bar 1 1 \bar 1]$, and $\hat\mathbf n_3 = 
[\bar 1\bar 1 1]$. Experimentally, a diamond optical lattice can be
generated by the interference of four laser beams with a suitable arrangement of light
polarizations \cite{toader2004}:
\[
V(\mathbf r) \propto \sum_{m=1}^3 \cos \left(\mathbf K_m \cdot \mathbf r\right)
-\cos \left( \mathbf K_0 \cdot \mathbf r\right).
\]
Here $\mathbf K_m = (\pi/2a)\,\hat\mathbf n_m$ is the laser wave vector, 
and $a$ is the nearest-neighbour bond length.
To obtain the orbital projectors on the nearest-neighbor bonds, we introduce a 
pseudovector $\bm\mu = (\mu^x, \mu^y, \mu^z) = (\lambda^{(6)}, 
\lambda^{(4)}, \lambda^{(1)})$ whose components are given by the 
three real-valued off-diagonal Gell-mann matrices.
The operators $\mu^a$ have the following nonzero elements:
$\langle p_y|\mu^x|p_z\rangle = \langle p_z|\mu^y|p_x\rangle 
= \langle p_x|\mu^z|p_y\rangle = 1$.
The orbital projectors along the four different bonds are
\begin{eqnarray}
P^{m} = ({I} + \sqrt{3}\, \bm\mu\cdot\hat\mathbf n_m )/3, 
\quad\quad (m = 0, 1, 2, 3).
\end{eqnarray}
Substituting the above expression into Eq.~(\ref{eq:Hex}) yields 
an effective Hamiltonian:
\begin{eqnarray}
\label{eq:H2}
H_{\rm diamond} = \frac{2J}{3} \sum_{m = 0}^3\,\sum_{\langle ij \rangle \parallel m}
\left(\bm\mu_i\cdot\hat\mathbf n_m\right)\left(\bm\mu_j\cdot\hat\mathbf
n_m\right).
\end{eqnarray}
Since the three matrices $\mu^a$ do not commute with each other, 
Eq.~(\ref{eq:H2}) defines a quantum `tetrahedral' Hamiltonian for pseudovectors 
$\bm\mu_i$ on the diamond lattice.
The exchange interaction (\ref{eq:H2}) is geometrically frustrated.
To see this, consider a bond $\langle ij \rangle$ along $[111]$ direction.
Its energy is minimized by orbital states $|\psi_i\rangle = |p_x + p_y + p_z\rangle/\sqrt{3}$ 
and $|\psi_j\rangle = |p_x - p_y\rangle/\sqrt{2}$. 
The corresponding expectation values of the pseudovector are 
$\langle \bm\mu_i \rangle = 2\,\hat\mathbf n_0/\sqrt{3}$ 
and $\langle \bm\mu_j \rangle = -\hat\mathbf z$, respectively. 
However, such an antiferro-orbital alignment can not be achieved simultaneously
on the other three $\langle 111 \rangle$ bonds attached to site $i$.

\begin{figure}[t]
\includegraphics[angle=0,width=0.9\columnwidth]{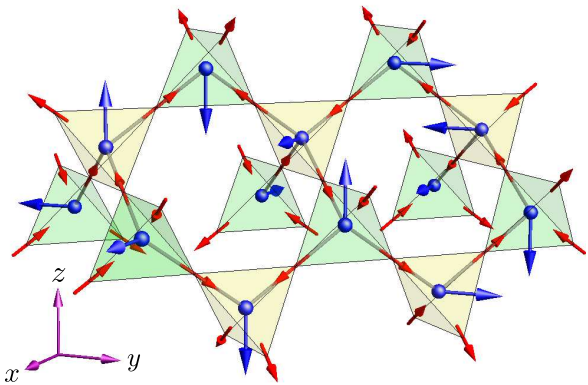}
\caption{A configuration of the pseudovectors on the diamond lattice
and its mapping to the spin-ice state on the medial pyrochlore lattice. 
The pseudovector only assumes six different values $\langle\bm\mu_i\rangle 
= \pm\hat\mathbf x$, $\pm\hat\mathbf y$, and $\pm\hat\mathbf z$ in the 
ground states, corresponding to $(p_y \pm p_z)$, $(p_z\pm p_x)$, and $(p_x\pm p_y)$
orbitals, respectively. These six orbital configurations are mapped to
the six 2-in-2-out ice states on a tetrahedron [Eq.~(\ref{eq:ice-spin})].}
\label{fig:ice-mapping}
\end{figure}

In order to understand the ground-state structure, we first minimize the 
Hamiltonian using the Gutzwiller ansatz:  
\begin{equation}
|\Psi\rangle = \prod_i |\psi_i\rangle =  \prod_{i}|\theta_i,\phi_i\rangle.
\end{equation}
The Gutzwiller wavefunction is a direct product of single-site orbitals,
The orbital wavefunction at a given site is parameterized by two angles 
$\theta$ and $\phi$: 
\[
|\psi\rangle = \sin\theta\cos\phi |p_x\rangle 
+ \sin\theta\sin\phi |p_y\rangle + \cos\theta |p_z\rangle.
\]
The expectation value of the pseudovector~is 
\begin{equation}
\label{eq:mu}
\langle \bm\mu \rangle = \left(\sin2\theta \sin\phi, 
\,\,\sin2\theta \cos\phi,\,\, \sin^2\theta \sin2\phi\right).
\end{equation}
Fig.~\ref{fig:domain2}(b) shows the domain of $\langle \bm\mu \rangle$ which  
has a tetrahedral symmetry.
We employ the Monte~Carlo simulations to minimize the resulting mean-field 
energy $ E\left\{\langle \bm\mu_i \rangle\right\} = \langle \Psi| 
H_{\rm diamond} |\Psi\rangle$, which is a function of the pseudovectors. 
Specifically, small changes of $\theta_i$ and $\phi_i$ are generated 
randomly and Eq.~(\ref{eq:mu}) is used to compute the change in 
$\langle\bm\mu_i\rangle$ and the corresponding $\Delta E$. 
These updates are then accepted according to detailed balancing.
The Monte Carlo minimization yields many degenerate Gutzwiller ground states. 
We find that the pseudovectors in the ground states point along one of 
the six cubic directions, {\it i.e.},
$\langle \bm\mu_i \rangle = \pm \hat\mathbf x$, $\pm\hat\mathbf y$, 
or, $\pm\hat\mathbf z$ for all sites [Fig.~\ref{fig:ice-mapping}], reminiscent
of the six-vertex model.
The corresponding orbital wavefunctions are $|\!\pm\!\hat\mathbf x\rangle 
= |p_y \pm p_z\rangle/\sqrt{2}$, and so on. 
The energy of each bond is exactly $\epsilon = -2J/9$ in the ground state.

Remarkably, the Gutzwiller ground states are also exact eigenstates of the Hamiltonian~(\ref{eq:H2}).
To see this, we define an Ising variable for each of the nearest-neighbor 
bonds $m$ attached to site $i$:
\begin{eqnarray}
\label{eq:ising}
\sigma^m_i = \sqrt{3}\,\langle\bm\mu_i\rangle\cdot\hat\mathbf 
n_m = \pm 1, \quad\quad (m = 0, 1, 2, 3).
\end{eqnarray}
They satisfy the orbital `ice rules':
\begin{eqnarray}
	\sigma^m_i\,\sigma^m_j = -1
\end{eqnarray}
for all nearest neighbors $\langle ij \rangle$ in the ground state.
Now consider a given site $i$, if the Ising variable $\sigma^m_i = -1$ 
on $m$-th bond, $|\psi_i\rangle$ is an eigenstate of the
operator $\bm\mu_i\cdot\hat\mathbf n_m$ with eigenvalue $-1/\sqrt{3}$.
On the other hand, for bonds with $\sigma^m_i = +1$, an extra term 
is generated when acted by the same operator. 
Specifically, let $|\psi_i\rangle = |\!+\!\hat\mathbf x\rangle$.
The Ising variable is positive on $[111]$ and $[1\bar1 \bar1]$ bonds; we have 
\begin{eqnarray*}
\left(\bm\mu_i\cdot\hat\mathbf n_m\right)\,|\psi_i\rangle 
= \pm\sqrt{2/3}\,|p_x\rangle +\sqrt{1/3}\,|\psi_i\rangle,
\end{eqnarray*}
with $\pm$ sign corresponding to $m=0$ and 1, respectively. 
Applying the combined bond operator on the Gutzwiller wavefunction yields
\begin{eqnarray*}
\left(\bm\mu_i\cdot\hat\mathbf n_m\right)\left(\bm\mu_j\cdot\hat\mathbf 
n_m\right) |\Psi\rangle= \mp \sqrt{2}/3\, 
|p_x\rangle_i\otimes|\tilde\Psi_i\rangle - 1/3\,|\Psi\rangle,
\end{eqnarray*}
where $|\tilde\Psi_i\rangle \equiv \prod_{k\neq i} |\psi_k\rangle$.
Note that the nearest-neighbor site $j = j(m)$ depends on the bond index $m$. 
The two extra terms with opposite signs cancel each other when summed
over $m=0$ and 1. 
The Gutzwiller state $|\Psi\rangle$ is thus an eigenstate of the sum 
of the two bond operators with positive $\sigma_i^m$.
Similar results hold for $|\psi_i\rangle = | \pm \hat\mathbf y\rangle$ 
or $| \pm \hat\mathbf z\rangle$. 
Since each site has two bonds with $\sigma^m_i = +1$ attached to it,
the extra terms cancel out when summed over all bonds. Consequently,
the Gutzwiller state $|\Psi\rangle$ is an exact eigenstate of the full Hamiltonian. 
We also performed exact diagonalization of Eq.~(\ref{eq:H2}) on a finite
system of 8 sites. 
With periodic boundary conditions, we find a huge degeneracy of the 
ground states which are indeed described by the Gutzwiller product.

We now employ the fact that pyrochlore is the {\it medial} lattice of diamond
to examine the degeneracy and structure of the quantum ground states. 
As shown in Fig.~\ref{fig:ice-mapping},
a pyrochlore magnet can be constructed by placing spins at the bond 
midpoints of a diamond lattice. 
This construction allows us to map the pseudovector field 
$\langle \bm\mu_i \rangle$ to a spin ice state on the pyrochlore lattice. 
Specifically, we label spins on a pyrochlore lattice by bond 
index $\langle ij \rangle$ of the diamond lattice and use the 
Ising variables (\ref{eq:ising}) to define its direction:
\begin{eqnarray}
\label{eq:ice-spin}
\mathbf S_{\langle ij \rangle} = +\sigma^m_i\,\hat\mathbf n_m 
= -\sigma^m_j\,\hat\mathbf n_m.
\end{eqnarray}
Here $\hat\mathbf n_m$ is a unit vector pointing from sites $i$ to $j$. 
Note that the diamond-lattice sites are located at centers of tetrahedra 
in the pyrochlore lattice, the above mapping shows that the six 
distinct values of pseudovectors in the ground state, i.e. 
$\langle\bm\mu_i\rangle = \pm\hat\mathbf x$, $\pm\hat\mathbf y$, 
and $\pm\hat\mathbf z$, correspond to the six different 2-in-2-out 
ice states on a tetrahedron as demonstrated in Fig.~\ref{fig:ice-mapping}.
The ground-state degeneracy of the diamond orbital model can thus be 
calculated using the so-called Pauling estimate which gives a residual 
entropy per site $s_0 \approx k_B \ln 3/2 \approx 0.405\, k_B$.

\begin{figure}[t]
\includegraphics[width=0.95\columnwidth]{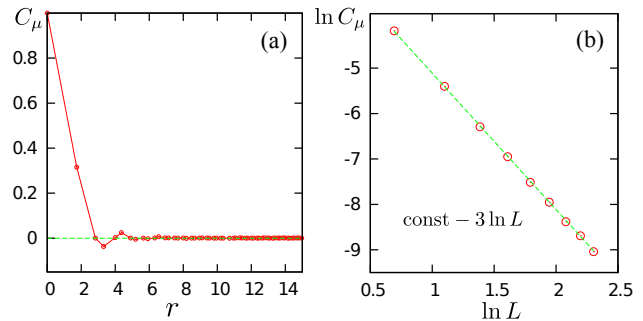}
\caption{(a) Orbital correlation function $C_{\mu}(r) 
= \langle \bm\mu(r)\cdot\bm\mu(0)\rangle$
as a function of distance $r$ in the quantum ground state of 
Hamiltonian~(\ref{eq:H2}). (b) shows $\ln C_{\mu}(L/2)$
as a function of $\ln L$, here $C_{\mu}(L/2)$ is the correlation 
function between sites separated by half the
linear size $L$ along a $\langle 110 \rangle$ chain of the lattice. 
The linear dependence in the log-log plot indicates a power-law decay: $C_{\mu}(L/2)\sim L^{-3}$.}
\label{fig:mc-ice}
\end{figure}

The above mapping also makes it possible to compute orbital correlation
functions by performing classical Monte Carlo simulations on pyrochlore 
spin ice. Since single-spin flip violates the ice rules, here we use the
non-local loop moves to navigate the manifold of spin-ice ground states \cite{barkema,melko}; 
the results are shown in Fig.~\ref{fig:mc-ice}.
The correlation function $C_{\mu}(r)=\langle\bm\mu(r)
\cdot\bm\mu(0)\rangle$ decays rather rapidly with the separation of spins. 
It is interesting to note that the pseudovector is related to the divergence-free
flux via $\mathbf B(\mathbf r_i) \sim \pm \langle \bm\mu_i \rangle$, where $\pm$ 
sign refers to the two sublattices of the diamond lattice. 
As discussed in the introduction, the magnetic field $\mathbf B$, hence
the pseudovectors, display a dipolar-like correlation function at long distances, 
as confirmed by our Monte Carlo simulations [Fig.~\ref{fig:mc-ice}(b)].  

\section{Summary and discussion}

To summarize, we have investigated the orbital exchange physics of $p$-band
spinless fermions on both cubic and diamond lattices. In both cases we have
found a macroscopic ground state degeneracy. The frustrated orbital interaction 
on the cubic lattice is governed by a classical three-state anisotropic Potts model. 
The ground state retains a finite entropy density $s_0 \approx 0.599 k_B$ per site.
Orbital correlation function decays exponentially at large distances. 
We have also derived a novel quantum `tetrahedral' model describing orbital 
interactions on the diamond lattice. We have obtained exact quantum many-body 
ground states which are extensively degenerate with a residual entropy 
density $s_0 \approx k_B \ln 3/2 \approx 0.405\, k_B$. By mapping the degenerate
quantum ground states to spin-ice states on a pyrochlore lattice, we have shown that
the fermionic $p$-band Mott insulators on a diamond lattice can 
be viewed as an orbital analog of the frustrated ice phase $I_c$ of water.

The huge degeneracy of orbital ice also helps circumvent the entropy obstacle
in its experimental realization. As noted in Ref.~\cite{ho}, a major challenge
in creating strongly correlated phases in cold-atom systems is reaching the
low level of entropies in such states. In this respect, the macroscopic residual 
entropy of the orbital ice renders the Coulomb phase much easier to realize in cold-atom optical lattices.

It is worth noting that the orbital ice model presented in this paper is different in nature 
from most conventional ice systems. First, the fundamental degrees of freedom of orbital
ice are orbital triplets defined on the diamond lattice, whereas those of the
conventional ice models are Ising-like variables on pyrochlore. Second, the pyrochlore
ice models with the ice rules explicitly incorporated into the Hamiltonian are essentially
classical systems. On the other hand, the orbital ice is an intrinsic quantum model. 
The orbital `ice rules' are emergent phenomena resulting from the orbital exchange dynamics.
This is a rare example of emergent geometrical frustration in three dimensions.
As usually happens in highly frustrated systems, the huge orbital degeneracy renders 
the ice phase susceptible to nominally small perturbations.
Various interesting phases could emerge from the orbital Coulomb phase.
Finally, it is also of great interest to examine the elementary excitations of the orbital ice model.

{\it Acknowledgment.} GWC thanks insightful discussions with C. D. Batista
and the supported of ICAM and NSF Grant DMR-0844115.
CW acknowledges the support of NSF under DMR-1105945 
and AFOSR-YIP program. 


\end{document}